\documentclass[reprint,showpacs,preprintnumbers,amsmath,amssymb,prl]{revtex4-1}
\usepackage[hidelinks]{hyperref}
\usepackage{amsmath}
\usepackage{amsfonts}
\usepackage{amsthm}
\usepackage[toc,page]{appendix}
\usepackage{dsfont}
\usepackage{graphics}
\usepackage{graphicx,bm}
\usepackage[caption=false]{subfig}
\usepackage{amssymb}
\usepackage{mathrsfs}
\usepackage{braket}
\usepackage{float}
\usepackage{color}
\usepackage{url}
\usepackage[abs]{overpic}
\usepackage[usenames,dvipsnames]{xcolor}
\usepackage{commath}
\usepackage{subfig}
\usepackage{multirow}
\usepackage{verbatim}
\usepackage{physics}
\usepackage{array}
\graphicspath{{Images/}}
\newcommand{\ii}{\mathrm{i}}
\newcommand{\ee}{\mathrm{e}}

\definecolor{darkblue}{rgb}{0,0,0.93} 
\definecolor{darkred}{rgb}{0.8,0,0} 

\hypersetup{
colorlinks=false, 
}

\begin{document}

\title{Dirac quantum walks with conserved angular momentum}
\author{Gareth Jay$^{1}$}
\email{gareth.jay@uwa.edu.au}
\author{Pablo Arnault$^{2}$ $^{3}$}
\email{pablo.arnault@ific.uv.es }
\author{Fabrice Debbasch$^{4}$}
\email{fabrice.debbasch@gmail.com}
\affiliation{{$^{1}$Physics Department, The University of Western Australia, Perth, WA 6009, Australia}\\
{$^{2}$Departamento de Física Teórica and IFIC, Universidad de Valencia and CSIC, Dr. Moliner 50, 46100 Burjassot, Spain}\\
{$^{3}$Institute for Quantum Computing, 200 University Ave W, Waterloo, ON N2L 3G1, Canada}\\
{$^{4}$Sorbonne Universit\'e, Observatoire de Paris, Universit\'e PSL, CNRS, LERMA, F-75005, {\sl Paris}, France}}
\date{\today}
\begin{abstract}
A Quantum Walk (QW) simulating the flat $(1 + 2)$D Dirac Eq.\ on a spatial polar grid is constructed.
Because fermions are represented by spinors, which do not constitute a representation of the rotation group, but rather of its double cover, the QW can only be defined globally on an extended spacetime where the polar angle extends from $0$ to $4 \pi$.
The coupling of the QW with arbitrary electromagnetic fields is also presented.
Finally, the cylindrical relativistic Landau levels of the Dirac Eq.\ are computed explicitly and simulated by the QW. 

\end{abstract}

\maketitle

\section{Introduction}

First proposed by Feynman as possible discretizations of Dirac path integrals \cite{feynman2010quantum,schweber1986feynman}, Quantum Walks (QWs) are unitary quantum automata that can be viewed as formal generalizations of classical random walks.
Reintroduced later by Aharonov et al. \cite{aharonov1993quantum}, and then studied systematically by Meyer \cite{meyer1996quantum}, QWs, like classical random walks in classical computing, have found application in quantum information and algorithmic development \cite{ambainis2007quantum,magniez2011search,ManouchehriWang2014}.
They can also be used as quantum simulators \cite{Strauch2006, Strauch2007, Kurzynski2008, Chandrashekar2013, Shikano2013, Arrighi2014,arrighi2016quantum, Molfetta2014, perez2016asymptotic}, where the lattice represents a discretization of continuous space, that could potentially represent a realistic discrete spacetime underlying the apparently continuous physical universe \cite{bisio2016special}.

It has been shown that several discrete-time quantum walks defined on regular square lattices simulate the Dirac dynamics in various spacetime dimensions and that these Dirac Quantum Walks (DQWs) can be coupled to various discrete gauge fields \cite{di2013quantum, di2014quantum, arnault2016landau,arnault2016quantum, arnault2016quantum2,arnault2017quantum, bisio2015quantum, marquez2018electromagnetic, bialynicki1994weyl, cedzich2013propagation, cedzich2018quantum}.
Extensions to regular non-square lattices have also been proposed \cite{jay2019dirac,arrighi2018dirac,jay2018new}.

More recently, a discrete action principle has been constructed for quantum automata \cite{debbasch2019action}.
In this context, the charge current of 1D DQWs has been recovered and a stress-energy `tensor' for DQWs has been constructed.
In particular, a `true' Hamiltonian (as opposed to an effective Hamiltonian), and a linear momentum for $1$D DQWs have been proposed and their conservation has been established for free 1D QWs.
These results extended to QWs defined on higher dimensional square lattices.
It is however not obvious that angular momenta can be built for QWs.
This question is not purely of fundamental, but also of practical interest, to simulate problems with axial symmetry.
For example, the Landau levels of an electron in a constant uniform magnetic field are degenerated and the conserved angular momentum can be used to distinguish between states sharing the same energy.
An experimental proposal based on magnetic discrete-time quantum walks has been made to construct anomalous Floquet Chern  topological insulators, that exhibit edge charge currents similar to those observed in the quantum Hall effect \cite{Sajid2019}.

The aim of this article is to construct angular momentum for DQWs. In standard theories, angular momentum is conserved under rotations if the system has rotational symmetry. We want to obtain such a theorem in the simplest manner possible, so we define the DQW on a polar space grid, which by definition has a natural rotational symmetry. We then simulate the so-called cylindrical Landau levels.
The material is organized as follows.
We first transcribe the usual, flat-spacetime $(1 +2)$D Cartesian Dirac Equation (CDE) into a Polar Dirac Equation (PDE), then construct a DQW which simulates the PDE in the presence of an arbitrary electromagnetic field.
We exhibit the angular momentum of this DQW, establish its conservation in electromagnetic fields with axial symmetry and finally construct the discrete Landau levels and show by numerical simulation that these converge to the usual continuous-spacetime Landau levels as the step if the spacetime grid tends to zero.
These results are summarized and discussed in the final section.
Appendix \hyperref[AppendixA]{A} offers an alternative derivation of the PDE while Appendix \hyperref[AppendixB]{B} presents the literal computations behind the construction of the Landau levels.

%

\section{The Polar Dirac Equation}

In $(1+2)$D flat spacetime, the CDE can be written 
\begin{equation}\label{DiracCartesian}
   {\mathcal D}^A_B\Psi^B
   =0,
\end{equation}
with the operator $\mathcal D$ defined by
\begin{equation}
{\mathcal D} = \ii (\gamma^0 \partial_t + \gamma^1 \partial_x + \gamma^2 \partial_y) - m,
\end{equation}
where $(t, x, y)$ are Minkowski coordinates in the flat $(1+2)$D spacetime and $m$ is the mass of the particle.
The indices $(A, B) \in \{ L, R \}^2$  refer to components on a cartesian, point-independent spin basis which we denote by $\left(b_L, b_R\right)$.
In this basis, the $\gamma$ operators are represented by the Pauli matrices:
\begin{equation}
[(\gamma^0)^A_B] = \sigma_1,\quad [(\gamma^1)^A_B] = \ii\sigma_2,\quad [(\gamma^2)^A_B] = \ii\sigma_3,
\end{equation}
where the notation $[(\gamma^i)^A_B]$ represents the matrix formed by the components of the operator $\gamma^i$ in the basis $\left(b_L, b_R\right)$.
We also introduce, for further use, a metric $\lambda$ in spin space, defined by
\begin{equation}
\lambda_{AB} = 
\left\{ \begin{array}{ll}
1 & \text{if} \ A = B \\
0 & \text{otherwise}
\end{array} \right. .
\end{equation}

To obtain the PDE from the CDE, one must proceed in two steps. The first one consists in introducing the polar coordinates $(r, \theta)$ in the plane and use
the relations $x = r \cos \theta$, $y = r \sin \theta$, to express the partial derivatives $\partial_x$ and $\partial_y$ appearing in $\mathcal D$ in terms of 
$\partial_r$ and $\partial_\theta$. This delivers 

\begin{equation}
{\mathcal D} = \ii \gamma^0 \partial_t + \ii {\tilde \gamma}^1(\theta)\partial_r - \frac{\ii}{r} {\tilde \gamma}^2(\theta)\partial_\theta-m,
\end{equation}
with
\begin{equation}
{\tilde \gamma}^1 (\theta)= \gamma^1\cos\theta+\gamma^2\sin\theta,
\end{equation}
and
\begin{equation}
{\tilde \gamma}^2(\theta) = \gamma^1\sin\theta-\gamma^2\cos\theta.
\end{equation}

The second step consists in changing basis in spin space. The new, so-called polar basis $\left(b_-, b_+\right)$, is defined by
\begin{eqnarray}
    b_-&=&\cos\frac{\theta}{2} b_L - \ii\sin\frac{\theta}{2} b_R,\\
    b_+&=&-\ii\sin\frac{\theta}{2} b_L  + \cos\frac{\theta}{2}  b_R,
\end{eqnarray}
and we denote by $\mathcal{M}_a^B$ and $(\mathcal{M}^{-1})_A^b$ the change of basis matrices:
\begin{eqnarray}
\Psi^B & = & {\mathcal M (\theta)}_a^B \Psi^a, \nonumber\\
\Psi^b & = & ({\mathcal M}^{-1}(\theta))_A^b \Psi^A,
\end{eqnarray}
where
\begin{equation}
    \left[\mathcal{M}(\theta)^B_a\right]=\ee^{-\ii\frac{\theta}{2}\sigma_1},\qquad  \left[(\mathcal{M}^{-1}(\theta))^b_A\right]=\ee^{\ii\frac{\theta}{2}\sigma_1}.
\end{equation}
In the new basis, the components of the operators $\gamma^0$, ${\tilde \gamma}^1$ and ${\tilde \gamma}^2$, read
\begin{equation}
[(\gamma^0)^a_b] = \sigma_1,\qquad [(\tilde{\gamma}^1)^a_b]=\ii\sigma_2,\qquad [(\tilde{\gamma}^2)^a_b]=\ii\sigma_3,
\end{equation}
and components of the operator $\mathcal D$ are
\begin{multline}\label{DiracPolar}
{\mathcal D}^a_b=   \ii\left(\gamma^1\right)^a_b\partial_t + \ii\left({\tilde \gamma}^2(\theta)\right)^a_b\partial_r \\+ \frac{\ii}{r}\left(\left({\tilde \gamma}^3(\theta)\right)^a_b\partial_\theta+\frac{1}{2}\ \left({\tilde \gamma}^2(\theta)\right)^a_b\right)-m.
\end{multline}
%
Note that the operators $\gamma^i$ and $\tilde{\gamma}^i$ are represented by the same matrices, but in different bases. 
Note also that the change of basis, being unitary, conserves the components of the metric $\lambda$ {\sl i.e.}
\begin{equation}
\lambda_{ab} = 
\left\{ \begin{array}{ll}
1 & \text{if} \ a = b \\
0 & \text{otherwise}
\end{array} \right. .
\end{equation}

In this new basis, the flat-spacetime Dirac equation reads ${\mathcal D}^a_b \Psi^b = 0$. This can be abbreviated into 
\begin{equation}
\left(\ii\sigma_1\partial_t - \sigma_2 \partial_r - \frac{1}{r}\left(\sigma_3 \partial_\theta+\frac{1}{2} \sigma_2\right)-m\right) \Psi = 0,
\end{equation}
which we call the Polar Dirac equation (PDE). We will use this compact form in the remainder of this article when no confusion with the CDE seems possible.

As usual, the coupling of the Dirac fermion with an electromagnetic field with $3$-potential $(A_\mu)=(A_t,A_r,A_\theta)$ is achieved by adding $+\ii q A_\mu$ to $\partial_\mu$ for $\mu = 0, 1, 2$. We choose to set the charge $q$ to $-1$ and get:
\begin{multline}
 \Bigg(\ii\sigma_1( \partial_t - \ii A_t) -\sigma_2(\partial_r - \ii A_r)  \\  -\frac{1}{r}\left(\sigma_3(\partial_\theta - \ii A_\theta)+\frac{1}{2}\sigma_2\right)-m\Bigg) \Psi = 0.
\end{multline}

%

Let us conclude this section by pointing out a very important property of the PDE. The second polar coordinate $\theta$ is an angle. Thus, the components $\Psi^L$ and $\Psi^R$ of $\Psi$ in the Cartesian spin basis, when written as functions of $r$ and $\theta$, are $2 \pi$-periodic functions of $\theta$. So are the time component $A_t$, the cartesian components $A_x$, $A_y$ and the polar components $A_r$ and $A_\theta$ of the potential. The components $\Psi^-$ and $\Psi^+$ of $\Psi$ in the polar spin basis are linear combinations of $\Psi^L$ and $\Psi^R$ with coefficients $\cos (\theta/2)$ and $\sin (\theta/2)$. These two coefficients are 
$2 \pi$-\emph{anti}-periodic in $\theta$ {\sl i.e.} they obey$f (\theta + 2 \pi) = - f(\theta)$ for all $\theta \in [0, 2 \pi[$. It follows that the polar components $\Psi^-$ and $\Psi^+$ are also $2 \pi$-anti-periodic in $\theta$. This expresses the fact that spinors belong to representations of the double cover of the rotation group $\mbox{SO}(2, {\mathbb R})$ and, thus get an extra minus sign after a rotation by $2 \pi$. Thus, the PDE is defined over $\{(r, \theta), r \in {\mathbb R}^*_+, \theta \in [0, 4 \pi [ \}$ and should only be used with initial conditions which are $2 \pi$-anti-periodic in $\theta$. By construction, the PDE conserves this anti-periodicity over time. 
Finally, only half integer modes $k = p + 1/2$, $p \in \mathbb Z$ enter the decomposition of the polar spinor components $\Psi^\pm$ in terms of Fourier modes $\exp (\ii k \theta)$.

Another method leading to the PDE is to use the so-called curved spacetime Dirac equation, which is actually valid for any spacetime, flat or curved, and in arbitrary coordinates, and particularize the treatment to $(1+2)$D flat Minkowski spacetime equipped with polar coordinates in $2$D physical space. This derivation is presented in Appendix \hyperref[AppendixA]{A}. The presentation retained above in the main part of this article has three distinct advantages: it is computationally simpler, it requires less geometry, and it highlights the global $2 \pi$-anti-periodicity of spinor components in the polar basis, which is not readily apparent from the purely local derivation given in Appendix \hyperref[AppendixA]{A}. 

However, things are different for quantum automata. There is indeed no way to obtain from a standard DQW defined on a cartesian grid a DQW approximating the PDE. For DQWs, the easiest route is to adapt the procedure presented in Appendix \hyperref[AppendixA]{A} for the Dirac equations. This is done is the next section.

\section{A polar Dirac Quantum Walk}

The starting point is the general construction presented in \cite{arnault2017quantum}, which delivers DQWs approximating Dirac equation in a possibly curved $(1 + 2)$D spacetime. By particularizing to flat Minkowski spacetime with $2$D polar coordinates, we will obtain a DQW which simulates the PDE, albeit without electromagnetic field. 
This field will be added as the latest step in the construction of the DQW.

\subsection{Without electromagnetic field}


The DQW will be defined on a grid in $(1+2)$D spacetime with temporal steps of $2\epsilon$ labelled by $\mathfrak{t}\in\mathbb{N}$, and identical spatial steps of $\epsilon$ labelled by $(\mathfrak{r},\mathfrak{h})\in\mathbb{N}\times[0,\frac{4\pi}{\epsilon}-1]$. The
DQW in Arnault et al. has the form 
\begin{equation}
\Phi_{\mathfrak{t}+1,\mathfrak{r},\mathfrak{h}}=\hat{V}\Phi_{\mathfrak{t},\mathfrak{r},\mathfrak{h}},   \label{eq:evol}
\end{equation}
where $\Phi_{\mathfrak{t},\mathfrak{r},\mathfrak{h}}=(\varphi^-_{\mathfrak{t},\mathfrak{r},\mathfrak{h}},\varphi^+_{\mathfrak{t},\mathfrak{r},\mathfrak{h}})^\top$ is a two component wave-function. The operator $\hat{V}$ reads
\begin{multline}\label{PDQW}
    \hat{V}=\Pi^{-1}\left[W_1(\alpha^{12})W_2(\alpha^{22})\right]\Pi\\\times\left[W_2(\alpha^{21})W_1(\alpha^{11})\right] Q(m\epsilon),
\end{multline}
where
\begin{equation}
    \Pi=\frac{1}{\sqrt{2}}\mqty(-\ii&1\\-1&\ii),
\end{equation}
the $W$ operators are defined as
\begin{equation}
    W_i(\alpha)=R^{-1}(\alpha)\left[U(\alpha)\hat{S}_iU(\alpha)\hat{S}_i\right]R(\alpha),
\end{equation}
the $\hat{S}$ operators are shift operators defined as
\begin{eqnarray}
    \hat{S}_1\Phi_{\mathfrak{t},\mathfrak{r},\mathfrak{h}}&=&\mqty(\varphi^-_{\mathfrak{t},\mathfrak{r}+1,\mathfrak{h}}\\\varphi^+_{\mathfrak{t},\mathfrak{r}-1,\mathfrak{h}}),\\
    \hat{S}_2\Phi_{\mathfrak{t},\mathfrak{r},\mathfrak{h}}&=&\mqty(\varphi^-_{\mathfrak{t},\mathfrak{r},\mathfrak{h}+1}\\\varphi^+_{\mathfrak{t},\mathfrak{r},\mathfrak{h}-1}),
\end{eqnarray}
and $U$ and $R$ are defined as
\begin{eqnarray}
    U(\alpha)&=&\mqty(-\cos\alpha&\ii\sin\alpha\\-\ii\sin\alpha&\cos\alpha),\nonumber\\
    R(\alpha)&=&\mqty(\ii\cos\left(\frac{\alpha}{2}\right)&\ii\sin\left(\frac{\alpha}{2}\right)\\-\sin\left(\frac{\alpha}{2}\right)&\cos\left(\frac{\alpha}{2}\right)).
\end{eqnarray}
The operator $Q$ is defined as
\begin{equation}
    Q(M)=\mqty(\cos(2M)&-\ii\sin(2M)\\-\ii\sin(2M)&\cos(2M)).
\end{equation}
The $\cos\alpha^{kl}$ terms match up the $n$-bein components like so:
\begin{equation}
    (e^\mu_a)=\mqty(e^t_0&e^t_1&e^t_2\\e^r_0&e^r_1&e^r_2\\e^\theta_0&e^\theta_1&e^\theta_2)=\mqty(1&0&0\\0&\cos\alpha^{11}&\cos\alpha^{12}\\0&\cos\alpha^{21}&\cos\alpha^{22}).
\end{equation}
The $n$-bein components are related to the metric components  $g_{\mu\nu}$ by
\begin{eqnarray}
    g^{\mu\nu}=e^\mu_ae^\nu_b\eta^{ab},
\end{eqnarray}
where 
$\eta_{ab}$ are the orthonormal components of the flat Minkowski metric. In the case of polar coordinates, one obtains:
\begin{equation}
  (e^\mu_a) =   \mqty(\dmat[0]{1,1,\frac{1}{r}}),
\end{equation}
which defines the four angles for the walk as:
\begin{eqnarray}
    \alpha^{11}&=0,\\
    \alpha^{12}=\alpha^{21}&=\frac{\pi}{2},\\
    \alpha^{22}&=&\arccos(\frac{1}{r}).
\end{eqnarray}


\subsection{With electromagnetic field}

As is the case with other DQWs, an electromagnetic field can be inserted by multiplying the advancement operator $\hat{V}$ at each point by an additional unitary operator $U^{\mbox{\small em}}$. The method presented in \cite{jay2018new} delivers
\begin{equation}\label{Uem}
U^{\mbox{\small em}} = \ee^{2\ii\epsilon A_t}\mqty(\ee^{-2\ii\epsilon A_r}&0\\0&\ee^{2\ii\epsilon A_r})\mqty(\cos\left(\frac{2}{r}\epsilon A_\theta\right)&\sin\left(\frac{2}{r}\epsilon  A_\theta\right)\\-\sin\left(\frac{2}{r}\epsilon  A_\theta\right)&\cos\left(\frac{2}{r}\epsilon  A_\theta\right))\nonumber\\.
\end{equation}

As the PDE, this walk only makes sense if the initial condition contains only half-integer Fourier modes. It can be checked by a direct computation that the walk then does not populate integer Fourier modes {\sl i.e.} that the two polar components of the walk wave-function then remain at all time $2 \pi$-anti-periodic functions of the angle $\theta$.

\subsection{Continuum Limit}
To obtain the continuum limit we take the same approach as in \cite{di2013quantum,di2014quantum,arnault2016landau,arnault2016quantum,arnault2016quantum2,arnault2017quantum,jay2019dirac,jay2018new} where we interpret $\Phi_{\mathfrak{t},\mathfrak{r},\mathfrak{h}}$ and the $\alpha^{ij}_{\mathfrak{t},\mathfrak{r},\mathfrak{h}}$ angles as functions $\Phi$ and $\alpha^{ij}$ at the polar spacetime coordinates of $(t=2\mathfrak{t}\epsilon, r=\mathfrak{r}\epsilon,\theta=\mathfrak{h}\epsilon)$. The factor of two on the temporal steps was established as necessary in \cite{arnault2017quantum} to make the continuum match with the standard form of the (curved spacetime) Dirac Equation. The limit of $\epsilon\rightarrow 0$ is then determined by Taylor expanding to first order in $\epsilon$. While the zeroth-order terms cancel each other out, the first order coefficients deliver the equation

\begin{multline}
    \bigg(\ii\sigma_1\left(\partial_t-\ii A_t\right)-\sigma_2\left(\partial_r-\ii A_r\right)\\-\frac{1}{r}\sigma_3\left(\partial_\theta-\ii A_\theta\right)-m\bigg)\Phi=0,
\end{multline}
which transcribes into the PDE for the wave-function $\Psi (t, r, \theta)=\frac{1}{\sqrt{r}}\Phi(t, r, \theta)$.

\section{Angular momentum}

Working  with the PDE and the PDQW pays off when one has to deal with angular momentum. Consider for example the PDE with a potential $A$ which does not depend on $\theta$. Writing Eq.\ \eqref{eq:evol} in $\theta$ Fourier space shows immediately that all wave numbers are decoupled and evolve unitarily independently of each other. This implies that the
average wave-number is conserved and this average coincides with the average of the operator ${\hat J} = - \ii \partial_\theta$, the sign has been chosen for reasons which will soon be made clear. This operator represents the total angular momentum of the Dirac field. The same {\sl mutatis mutandi} goes for the PDQW, so that ${\hat J}$ can also be considered/defined as the angular momentum of the PDQW, the main difference being that
Fourier analysis now takes place on a bounded grid, so the spectrum is also discrete and bounded.

Let us now show that ${\hat J}$ can be interpreted as the sum of the orbital angular momentum and of the spin of the Dirac field. Indeed, one has (with obvious notations)
\begin{eqnarray}
\expval{\hat{J}}&=&- \ii  \int \lambda_{ab} \Psi^{a *} \frac{\partial \Psi^b}{\partial\theta}  r \dd r \dd\theta \nonumber \\
& = & -\ii  \int \Psi^*_{b}  ( {\mathcal M}^{*}(\theta))_D^b {\mathcal M (\theta)}_c^D\frac{\partial  \Psi^c}{\partial\theta}  r \dd r \dd\theta \nonumber \\
& = &  -\ii \int \Psi^*_{D}  \left[ \frac{\partial}{\partial\theta} \left({\mathcal M (\theta)}_c^D \Psi^c \right) - 
\frac{d{\mathcal M (\theta)}_c^D}{d\theta} \Psi^c
 \right] r \dd r \dd\theta.\nonumber\\
\end{eqnarray}
This expression can be further simplified in the following manner. First, 
\begin{equation}
{\mathcal M (\theta)}_c^D \Psi^c = \Psi^D,
\end{equation}
so the first partial derivative with respect to $\theta$ on the right-hand side of the last equation should actually be expressed in terms of partial derivatives with respect to $x$ and $y$. A simple computation shows that $\partial_\theta = x \partial_y - y \partial_x$. Second, computing the derivative of $\mathcal M$ with respect to $\theta$ delivers

\begin{equation}
\frac{d{\mathcal M (\theta)}_c^D}{d\theta} = - \frac{\ii}{2} (\sigma_1)^D_E {\mathcal M (\theta)}_c^E.
\end{equation}
Putting everything together leads to
\begin{eqnarray}
\expval{\hat{J}}&=&- \ii  \int \lambda_{AB} \Psi^{A *} \left( 
(x \partial_y - y \partial_x)\delta^B_C \right. \nonumber \\
& & \left. + \frac{\ii}{2} (\sigma_1)^B_C
\right) \Psi^C r \dd r \dd\theta.
\end{eqnarray}
The first term on the right-hand side represents the kinetic angular momentum and the second represents the spin.

\section{Quantum simulation of relativistic Landau levels}

Relativistic Landau levels are eigenstates of the Dirac Hamiltonian in the presence of a uniform magnetic field orthogonal to the plane of motion. These levels are degenerate and any operator which commutes with the Hamiltonian can be used to label the different eigenstates corresponding to the same level. Because the magnetic field is
uniform and orthogonal to the plane of motion, the angular momentum operator commutes with the Hamiltonian and can be used to distinguish between eigenstates of a given Landau level, and we thus search for eigenstates of the form $\Phi_{E, \kappa}(t, r, \theta) = \exp(- i E t)\Xi (r) \exp(- i \kappa \theta)$. The computation of these eigenstates is best carried out by replacing the components $\xi^-$ and $\xi^+$ of $\Xi$ by the new unknown functions
\begin{eqnarray}
u^- & = & \frac{\ii}{\sqrt{2}} \exp(\frac{\ii \pi}{4}) ( \xi^- + \xi^+), \nonumber\\
u^+ & = & \frac{1}{\sqrt{2}} \exp(\frac{\ii \pi}{4}) (-  \xi^- + \xi^+).
\end{eqnarray}
%
%
%
%
%
%
%
The eigenfunctions of energy $E$ and angular momentum $\kappa$ then obey

\begin{multline}\label{DEtoSolve}
     \pm\partial_ru^\pm_{E,\kappa}(r)+\left(\frac{\kappa}{r}+\frac{1}{2}Bqr\right)u_{E,\kappa}^\pm(r)-(E\mp m)u_{E,\kappa}^\mp(r)=0.
\end{multline}
This explicit solution of this system is presented in Appendix \hyperref[AppendixB]{B}. For example, if $Bq>0$, one obtains
\begin{eqnarray}
    u_{E,\kappa}^-(r)&=&\frac{Bq}{m-E}Cr^{1-\kappa}\ee^{-\frac{1}{4}Bqr^2}L_{n-1}^{\alpha+1}\left(\frac{1}{2}Bqr^2\right),\nonumber\\
    u_{E,\kappa}^+(r)&=&Cr^{-\kappa}\ee^{-\frac{1}{4}Bqr^2}L^a_n\left(\frac{1}{2}Bqr^2\right),
\end{eqnarray}
where $L_n^\alpha(x)$ are associated Laguerre polynomials, $n\ge1$ and $\alpha\ge-n$ are integers that are related to energy, mass and angular momentum as
\begin{eqnarray}
    n&=&\frac{E^2-m^2}{2Bq},\\
    \alpha&=&-\kappa-\frac{1}{2},
\end{eqnarray}
and the constants $C$ is defined by the normalisation condition:
\begin{equation}
     \abs{C}^2=\frac{(m-E)^2(Bq)^{\alpha+1}n!}{\pi2^{\alpha+1}(n+\alpha)!(2Bqn+(m-E)^2)}.
\end{equation}

\begin{figure}
    \centering
    \includegraphics[width=0.5\textwidth]{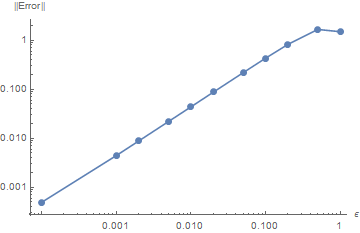}
    \caption{Evolution of the discretisation error $\delta$ with the time and space step $\epsilon$ for parameters $n=1$, $\alpha=5$, $Bq=0.1$ and $m=1$}
    \label{fig:errorPlot}
\end{figure}

Choose an eigenfunction of energy $E$ and angular momentum $\kappa$, say $\Phi_{E, \kappa}$, and use it as initial condition for  the polar DQW with finite discretisation parameter $\epsilon$. After one time-step of length $\epsilon$, the
{\sl continuous} dynamics of the Dirac equation changes the function simply by the phase factor $\exp(- \ii E \epsilon)$ while the discrete dynamics of the walk delivers a different function $\Phi^1_{E, \kappa}$. The error $\delta(\epsilon)$ involved in the discretisation can be measured by the $L^1$ norm of the difference $\Phi^1_{E, \kappa} - \exp(- \ii E \epsilon) \Phi_{E, \kappa}$, and this error should naturally tend to zero with $\epsilon$. Figure \ref{fig:errorPlot} shows the typical evolution of this error with the parameter $\epsilon$.

\section{Conclusion}

We have presented a new DQW which can simulate the $(1 + 2)$D flat-spacetime Dirac equation on a spatial polar grid. Thanks to the polar grid, we have identified a quantity which we define as the angular momentum of the DQW, since it is conserved under rotations when the system has rotational symmetry (e.g., for a free DQW, but also for a DQW with electromagnetic potential if the latter has rotational symmetry). Because fermions are described by spinors, the PDQW can only be defined globally on an extended spacetime grid. We have also shown how the PDQW can be coupled to arbitrary electromagnetic fields and we have demonstrated that the PDQW can simulate relativistic Landau levels. 

Let us now conclude by mentioning a few possible extensions to this work. A first one would be to build $(1 + 3)$D DQWs on a spherical spatial grid and, more generally, on an elliptical spatial grid. The global and local discrete $U(1)$ gauge invariance associated to electromagnetism and charge conservation should also be investigated on such non-cartesian grids. The same should be carried out for arbitrary Yang-Mills fields and for gravitational fields as well. Finally, some of the material developed in \cite{debbasch2019action} for DQWs on $(1 + 1)$D cartesian grids only should be extended to more general grids. For example, can one introduce on general grids an action principle which involves spacetime coordinates and delivers the stress-energy momentum of the walk?
%

\bibliographystyle{apsrev4-1}
\bibliography{Biblio.bib}

\appendix
\section{A: (2+1)D Dirac Equation in Polar Coordinates}\label{AppendixA}
An easy way to derive the PDE is to start with the general, so-called curved spacetime formulation of the Dirac equation and then particularise the treatment to polar coordinates and to a polar spin basis. 

Let us first recall that the Dirac equation in flat $(1+2)$D Minkowski spacetime (in natural units $\hbar=c=1$), in cartesian coordinates and in a cartesian spin basis takes the form:
\begin{equation}\label{Dirac3D}
    (\ii\gamma^a\partial_a-m)\Psi=0,
\end{equation}
where $\gamma^a$ are related to the flat Minkowski metric by
\begin{equation}\label{flatGamma}
    \{\gamma^a,\gamma^b\}=2\mathbb{I}\eta^{ab},
\end{equation}
where
\begin{equation}\label{minkMetric}
    \eta^{ab}=\eta_{ab}=\mqty(\dmat[0]{1,-1,-1}),
\end{equation}


Formally speaking, the general formulation of the Dirac equation can be obtained
by replacing the $\gamma^a$ matrices with coordinate dependent matrices $\gamma^\mu$ and replacing the partial derivatives with covariant derivatives $\nabla_\mu$. The coordinate dependent gamma matrices are defined as
\begin{equation}\label{coDepgamma}
    \gamma^\mu=e^\mu_a\gamma^a,
\end{equation}
where the $e^\mu_a$'s are the components of the $n$-bein vectors $e_a$ in the coordinate basis in the tangent space {\sl i.e.} 
\begin{equation}\label{localbasisVector}
    e_a=e^\mu_a\partial_\mu.
\end{equation}
The basis $(\theta^a)$ dual to $(e_a)$ has components $e^a_\mu$ in the basis $(dx^\mu)$ of the cotangent:
\begin{equation}\label{dualbasisVector}
    \theta^a=e^a_\mu \dd x^\mu.
\end{equation}

By definition,
\begin{equation}\label{spacetimeInterval}
    \dd s^2=g_{\mu\nu}\dd x^\mu \dd x^\nu=\eta_{ab}\theta^a\theta^b,
\end{equation}
where the $g_{\mu \nu}$'s are the coordinate basis components of the metric. 
 Substituting Eq.\ \eqref{dualbasisVector} into Eq.\ \eqref{spacetimeInterval} we get
\begin{equation}
    g_{\mu\nu}=\eta_{ab}e^a_\mu e^b_\nu\label{metricVerbein}.
\end{equation}
Assuming there is no torsion and the connection is compatible with the metric, the covariant derivative of spinors is defined by
\begin{equation}\label{covariantDerivative}
    \nabla_\mu=\partial_\mu+\Gamma_\mu,
\end{equation}
where the connection coefficients $\Gamma_\mu$ read
\begin{equation}\label{affineConnection}
    \Gamma_\mu=\frac{1}{8}\omega_{\mu cd}[\gamma^c, \gamma^d].
\end{equation}
The quantities $\omega_{\mu cd}=\eta_{ca}\omega^a_{\mu d}$ are called the Ricci rotation coefficients. They can be computed 
from the Christoffel symbols  are are related to the Christoffel symbols $\Gamma^\nu_{\sigma\mu}$ by
\begin{eqnarray}\label{spinConnection}
    \omega^c_{\mu d}=e^c_\nu e^\sigma_d \Gamma^\nu_{\sigma\mu}+e^c_\nu\partial_\mu e^\nu_d,
\end{eqnarray}
and the Christoffel symbols can be computed from the metric components by
\begin{equation}\label{christoffel1}
    \Gamma^\nu_{\sigma\mu}=g^{\nu\rho}\Gamma_{\rho\sigma\mu},
\end{equation}
\begin{equation}\label{christoffel2}
    \Gamma_{\rho\sigma\mu}=\frac{1}{2}(\partial_\mu g_{\rho\sigma}+\partial_\sigma g_{\rho\mu}-\partial_\rho g_{\sigma\mu}).
\end{equation}

Now, let us focus on $(1 + 2)$D flat Minkowski spacetime, choose a Lorentz frae $(t, x, y)$ and introduce the new coordinate system $(t, r, \theta)$ where $(r, \theta)$ are polar coordinates in the $(x, y)$ plane. In these new coordinates, the metric and inverse components are:
\begin{eqnarray}
    (g_{\mu\nu})&=&\mqty(\dmat[0]{1,-1,-r^2}),\\ (g^{\mu\nu})=(g_{\mu\nu})^{-1}&=&\mqty(\dmat[0]{1,-1,-\frac{1}{r^2}}).
\end{eqnarray}
In these coordinates, the only non vanishing Christoffel symbols are:
\begin{eqnarray}
    \Gamma_{r\theta\theta}&=&\frac{1}{2}(\partial_\theta g_{r\theta}+\partial_\theta g_{r\theta}-\partial_r g_{\theta\theta})=r,\nonumber\\
    \Gamma_{\theta r\theta}&=&\frac{1}{2}(\partial_\theta g_{\theta r}+\partial_r g_{\theta\theta} - \partial_\theta g_{r\theta})=-r,\nonumber\\
    \Gamma_{\theta\theta r}&=&\frac{1}{2}(\partial_r g_{\theta\theta}+\partial_\theta g_{\theta r}-\partial_\theta g_{\theta r})=-r,
\end{eqnarray}
\begin{eqnarray}
    \Rightarrow\Gamma^r_{\theta\theta}&=&g^{rr}\Gamma_{r\theta\theta}=-r,\nonumber\\
    \Gamma^\theta_{r\theta}=\Gamma^\theta_{\theta r}&=&g^{\theta\theta}\Gamma_{\theta\theta r}=g^{\theta\theta}\Gamma_{\theta r\theta}=\frac{1}{r}.
\end{eqnarray}
Equation \eqref{spacetimeInterval} gives
\begin{equation}
    \dd s^2=\dd t^2-\dd r^2-r^2\dd \theta^2=(\theta^0)^2-(\theta^1)^2-(\theta^2)^2,
\end{equation}
which results in 
\begin{equation}
    \theta^0=\dd t,\qquad \theta^1=\dd r,\qquad\theta^2=r\dd \theta.
\end{equation}
Using Eq.\ \eqref{dualbasisVector} then delivers the following $n$-bein and inverse $n$-bein components
\begin{equation}
    (e^\mu_a)=(e^a_\mu)^{-1}=\mqty(\dmat[0]{1,1,r})^{-1}=\mqty(\dmat[0]{1,1,\frac{1}{r}}),
\end{equation}
where the upper index indicates the rows and the lower index indicates the columns.

The only non vanishing Ricci rotation coefficients are
\begin{eqnarray}
    \omega^1_{\theta 2}&=&e^1_r e^\theta_2\Gamma^r_{\theta \theta}=(1)(\frac{1}{r})r=-1,\nonumber\\
    \omega^2_{\theta 1}&=&e^2_\theta e^r_1\Gamma^\theta_{r\theta}=(r)(1)\frac{1}{r}=1.
\end{eqnarray}
This leads to
\begin{eqnarray}
    \omega_{\theta12}&=&\eta_{11}\omega^1_{\theta2}=1,\nonumber\\
    \omega_{\theta21}&=&\eta_{22}\omega^2_{\theta1}=-1,
\end{eqnarray}
and the only non vanishing spin connection coefficients is therefore
\begin{eqnarray}
    \Gamma_\theta&=&\frac{1}{8}\omega_{\theta 12}[\gamma^1, \gamma^2]+\frac{1}{8}\omega_{\theta 21}[\gamma^2, \gamma^1]\nonumber\\
    &=&\frac{1}{4}[\gamma^1,\gamma^2].
\end{eqnarray}
%

The PDE thus reduces to
\begin{equation}
    \left(\ii\gamma^0\partial_t+\ii\gamma^1\partial_r+\frac{\ii}{r}\gamma^2\left(\partial_\theta+\frac{1}{4}[\gamma^1,\gamma^2]\right)-m\right)\Psi=0.
\end{equation}
Choosing the representation  where $\gamma^0=\sigma_1$, $\gamma^1=\ii\sigma_2$ and $\gamma^2=\ii\sigma_3$ gives us
\begin{equation}
    \left(\ii\sigma_1\partial_t-\sigma_2\partial_r-\frac{1}{r}\left(\sigma_3\partial_\theta+\frac{1}{2}\sigma_2\right)-m\right)\Psi=0,
\end{equation}
which coincides with the PDE presented in Section 2.

This equation conserves the normalisation condition
\begin{equation}
    \int\sqrt{g}e^t_0\Psi^\dagger\Psi\dd^3x=1,
\end{equation}
where $g=\det(g_{\mu\nu})$. If one works with $(t, r, \theta)$ as coordinates, $g=r^2$ and this normalisation condition can be rewritten 
\begin{eqnarray}
    \int\Phi^\dagger\Phi\dd^3x=1,
\end{eqnarray}
with 
\begin{equation}
\Phi = \sqrt{r} \Psi.
\end{equation}
The function $\Phi$ obeys
\begin{multline}
    \Bigg(\ii\gamma^0\partial_t+\ii\gamma^1\partial_r\\+\frac{\ii}{r}\gamma^2\left(\partial_\theta+\frac{1}{4}[\gamma^1,\gamma^2]\right)-m\Bigg)\left(\frac{1}{\sqrt{r}}\Phi\right)=0,
\end{multline}
which becomes, in the above representation:
\begin{equation}
    \left(\ii\sigma_1\partial_t-\sigma_2\partial_r-\frac{1}{r}\sigma_3\partial_\theta-m\right)\Phi=0.\label{aimingFor}
\end{equation}
This equation coincides with the continuum limit of the free polar quantum walk introduced in this article. 
\\

\section{B: Laguerre solutions to the differential equations}\label{AppendixB}
To solve the differential equations in Eq.\ \eqref{DEtoSolve}, we first make the substitutions $u^\pm=r^{\mp \kappa}e^{\mp\frac{1}{4}Bq r^2}v^\pm$. 
Assuming $E\ne\pm m$, $v^\pm$ satisfy the equations
\begin{eqnarray}
    v^-&=&\frac{1}{E-m}f(r)\dv{v^+}{r},\label{DE1}\\
    v^+&=&\frac{1}{E+m}g(r)\dv{v^-}{r},\label{DE2}
\end{eqnarray}
where $f(r)=r^{-2\kappa}\ee^{-\frac{1}{2}Bq r^2}$ and $g(r)=-\frac{1}{f(r)}$. Now differentiate Eq.\ \eqref{DE1} and substitute $\dv{v^-}{r}$ into Eq.\ \eqref{DE2} to get
\begin{equation}
    \dv[2]{v^+}{r}+\frac{1}{f}\dv{f}{r}\dv{v^+}{r}+(E^2-m^2)v^+.
\end{equation}
Now we focus on the case $Bq>0$ and make the change of variable $x=\frac{1}{2}Bqr^2$ (for $Bq<0$, the useful change of variable reads $x=-\frac{1}{2}Bqr^2$). Simplifying this brings us to:
\begin{equation}
    x\dv[2]{v^+}{x}+\left(-\kappa+\frac{1}{2}-x\right)\dv{v^+}{x}+\frac{E^2-m^2}{2Bq}v^+=0.
\end{equation}
Introducing $n=\frac{E^2-m^2}{2Bq}$ and $\alpha=-\kappa-\frac{1}{2}$, this can then be rewritten as 
\begin{equation}
    x\dv[2]{v^+}{x}+(\alpha+1-x)\dv{v^+}{x}+nv^+=0.
\end{equation}
We thus write
\begin{equation}
    v^+=CL_n^\alpha(x)=CL_n^\alpha\left(\frac{1}{2}Bqr^2\right),
\end{equation}
where $C$ is a constant and $L^n_\alpha$ is an associated Laguerre polynomial.
Substituting this into Eq.\ \eqref{DE1} we then obtain
\begin{equation}
    v^-=-\frac{CBqr}{E-m}f(r)L_{n-1}^{\alpha+1}\left(\frac{1}{2}Bqr^2\right) 
\end{equation}
and 
\begin{equation}
    u(r)=\mqty(\frac{Bq}{m-E}Cr^{1-\kappa}\ee^{-\frac{1}{4}Bqr^2}L_{n-1}^{\alpha+1}\left(\frac{1}{2}Bqr^2\right)\\Cr^{-\kappa}\ee^{-\frac{1}{4}Bqr^2}L^a_n\left(\frac{1}{2}Bqr^2\right)).
\end{equation}

Enforcing the normalisation condition 

\begin{equation}
\int_0^{2\pi} \int_0^\infty \abs{ \Phi (r, \theta) }^2 \dd r \dd\theta = 1.
\end{equation}
delivers
\begin{equation}
    \abs{C}^2=\frac{(m-E)^2(Bq)^{\alpha+1}n!}{\pi2^{\alpha+1}(n+\alpha)!(2Bqn+(m-E)^2)}.
\end{equation}
\end{document}